# A scaling relation of vortex-induced rectification effects in a superconducting thin-film heterostructure

Yusuke Kobayashi[1], Junichi Shiogai[1,2,†], Tsutomu Nojima[3], and Jobu Matsuno[1,2]

[1]*Department of Physics, Osaka University, Toyonaka, Osaka 560-0043, Japan*

[2]*Division of Spintronics Research Network, Institute for Open and Transdisciplinary Research Initiatives, Osaka University, Suita, Osaka 565-0871, Japan*

[3]*Institute for Materials Research, Tohoku University, Sendai 980-8577, Japan*

[†]Corresponding author: junichi.shiogai.sci@osaka-u.ac.jp

**Abstract**

Supercurrent rectification, nonreciprocal response of superconducting properties sensitive to the polarity of bias and magnetic field, has attracted growing interest as an ideal diode. While the superconducting rectification effect is a consequence of the asymmetric vortex pinning, the mechanisms to develop its asymmetric potentials have been a subject of ongoing debate, mainly focusing on microscopic breaking of spatial inversion symmetry and macroscopic imbalance of the sample structure. Here, we report on comparative study of the superconducting diode effect and nonreciprocal resistance in a superconducting Fe(Se,Te)/FeTe heterostructure. In normal state, we observe finite nonreciprocal resistance as a hallmark of the spin-orbit interaction with structural inversion asymmetry. In the superconducting state, we find that the strongly enhanced nonreciprocal coefficient in transition regime is directly coupled to the superconducting diode efficiency through a universal scaling law, indicating the role of spin-momentum-locked state on the asymmetric pinning potential. Our findings, providing a unified picture of the superconducting rectification, pave the way for functionalizing superconducting diode devices.

**Main**

A superconducting thin-film heterostructure with structural inversion asymmetry is a promising platform for observing intriguing physical properties and spintronic functionalities[1-5]. One of the archetypal consequences of the broken inversion symmetry is the nonreciprocal response under the magnetic field[6-18], where the resistivity in one direction is larger than that in the opposite direction. The concept of such a rectification effect induced by broken inversion symmetry was originally examined in twisted and polar non-superconducting conductors known as magnetochiral anisotropy[19,20]. When the inversion symmetry is broken along the $z$-direction, the phenomenological expression of the sheet resistance $R_{xx}$ at low excitation current density $J$ under magnetic field $B$ is given by

$$R_{xx}(B, J) = R_{xx}^0[1 + \gamma(\mathbf{z} \times \mathbf{B}) \cdot \mathbf{J}] \quad \text{Eq. (1)}$$

where $R_{xx}^0$ is the Ohmic (linear) resistance, and the second-term in the bracket corresponds to the nonreciprocal (non-linear) contribution. A nonreciprocal coefficient $\gamma$ represents a strength of polarization, which has been used for qualitative analysis of the nonreciprocal resistance[8].

Especially, in the case of superconductors with the broken inversion symmetry, it has been reported that the value of $\gamma$ is drastically enhanced around superconducting transition temperature ($T_c$) as exemplified in two-dimensional (2D) superconductors confined in gated surface[12], heterointerfaces[14,15], exfoliated flakes of non-centrosymmetric layered 2D materials[17]. These nonreciprocal phenomena are further extended to superconducting critical parameters[16,17,21,22], in that the critical current in one direction $I_c^+$ is larger (or smaller) than that in the opposite direction $I_c^-$ depending on the polarity of ***B***. By analogy with the semiconductor *p-n* junction, the nonreciprocity in critical current is referred as the superconducting diode effect, where the diode efficiency $\eta$ is defined as $\eta = (I_c^+ - I_c^-)/(I_c^+ + I_c^-)$.

One of the proposed mechanisms for the supercurrent rectification effect in earlier works was the rachet motion of vortices caused by macroscopically asymmetric pinning potential, which was introduced artificially in trivial superconductors[23-27]. Recently, a variety of other origins of the rachet mechanism, such as microscopically asymmetric band structure caused by Rashba-type or Ising-type spin-orbit interaction typically in 2D superconductors[10,12,17,28-30] and asymmetric edge barrier in a superconducting microbridge[16,18,31], have been suggested, depending on the electronic and geometric condition of the samples of interest. In most cases, the nonreciprocity or the diode effect has been discussed by measuring either $\gamma$ in the transition regime or $\eta$ in the zero-resistance superconducting regime, respectively, in spite that both should be related parameters. For comprehensive understanding of the superconducting rectification effect in the broad sense, it is highly desired to clarify the correlation between $\gamma$ and $\eta$ with systematic measurements of resistive transition and the critical current in magnetic fields. However, such experiments have been limited probably because the broadness of resistive transition width generally impedes the magnitude of critical current density under magnetic fields.

Here, we studied the nonreciprocal superconducting transport properties in a Fe(Se,Te)/FeTe heterostructure (Fig. 1a), where the inversion symmetry is broken along the out-of-plane direction. Distinct features of Fe(Se,Te) are the increase of superconducting critical parameters[32,33,34,35,36,37,38], such as $T_c$, the upper critical magnetic field ($B_{c2}$), and the critical current density ($J_c$), by Te substitution to Se sites, as well as the enhanced spin-orbit interaction[39]. These characteristics make Fe(Se,Te) thin-film and its heterostructure be good candidates to examine the supercurrent rectification with the microscopic origin. Indeed, our Fe(Se,Te)/FeTe heterostructure shows the nonreciprocal resistance with the finite $\gamma$ value in the normal state, as a hallmark of the strong Rashba spin-orbit interaction, which is largely enhanced by a factor of $10^6$ in the superconducting state around $T_c$. Simultaneously, we observed the relatively large diode efficiency $\eta$ up to 15% at temperature much below $T_c$ when the in-plane magnetic field of as large as a several teslas ($B // y$) is applied perpendicular to the current ($J // x$) (Fig. 1a). Furthermore, by comparing temperature and magnetic-field dependences of $\eta$ and $\gamma$, we found a universal scaling between them, which can be understood comprehensively using the asymmetric pinning potential for vortices. Judging from the observation of nonreciprocal transport at normal state, the Rashba spin-orbit interaction is suggested as an origin of the asymmetric vortex dynamics. Our demonstration on the unified picture for superconducting rectification effect sheds light on perspectives for material selection and structural design toward highly efficient supercurrent rectification devices.

**Superconducting critical parameters and vortex phase diagram.**

Figure 1b shows temperature dependence of $R_{xx}$ for the Se/Fe(Se,Te)/FeTe heterostructure together with the 57.1-nm-thick FeTe single layer film. A clear superconducting transition with onset $T_c$ ($T_c^{on}$) and zero-resistance $T_c$ ($T_c^{zero}$) is exhibited in the Se/Fe(Se,Te)/FeTe heterostructure at 14.0 K and 11.6 K, respectively. Figure 1c shows the $J$-$B$-$T$ phase diagram in the superconducting Se/Fe(Se,Te)/FeTe heterostructure. We achieved a larger critical current density $J_c(B,T)$ and the upper critical field $B_{c2}(T)$ than the values reported in FeSe bulk single crystal[32,34] (black circles) owing to Te doping. The critical current density $J_c$ was estimated to be $3.9 \times 10^5$ Acm$^{-2}$ at 4.2 K, which corresponds to 2.9 % of the depairing current density of $J_d = 1.3 \times 10^7$ Acm$^{-2}$. We defined the $B_{c2}(T)$ line from the temperatures when $R_{xx}(T)$ intersects the half-value of the normal-state resistance under the magnetic field. A contour plot of normalized $R_{xx}(T,B)$ is shown on the basal plane of Fig. 1d to construct a vortex phase diagram. The broadening of $R_{xx}$ below $T_c^{on}$ by application of $B$ can be well reproduced by the thermally activated flux flow (TAFF) model[34,40]. The robustness of the superconductivity with large critical parameters together with the wide TAFF region in the Fe(Se,Te)/FeTe heterostructure allows the comprehensive study of superconducting diode effect and nonreciprocal resistance under a wide range of the current and magnetic field.

**Superconducting diode effect.**

To evaluate the superconducting diode effect, the current-voltage (*I-V*) characteristics were measured under magnetic field. As shown in Fig. 2a ($B = \pm 5$ T, $T = 4.2$ K), the nonreciprocal *I-V* characteristics apparently emerges. For the positive ***B*** (+5 T, black curve in Fig. 2a), the critical current in positive bias ($I_c^+$) is larger than that in a negative bias ($I_c^-$). Such a relationship is reversed for the negative ***B*** (−5 T, red curve in Fig. 2a). This rectification effect against the in-plane ***B*** indicates the broken inversion symmetry along the direction perpendicular to the plane. Note that this rectification effect was not observed in out-of-plane **B**, excluding the effect of different barriers against vortex motion at left and right edges of superconducting channel[16,31] or inhomogeneous pinning potential[41], as pointed in previous low-field experiments. Figure 2b shows the $I_c^+$ and $I_c^-$ values as well as their averaged value $I_c^0 = (I_c^+ + I_c^-)/2$ (black solid line) as a function of $B$ up to 15 T at 4.2 K. Here, $I_c$ is the critical current defined at $V = 1$ mV. The $I_c^+(B)$ and $I_c^-(B)$ curves indicate that the sign of the nonreciprocal contribution to $I_c(B)$ is reversed by the polarity change in $B$.

To clarify the nonreciprocal component of $I_c$, the $I_c(B)$ is analyzed in the form of $I_c^{+/-} = I_c^0 \pm \Delta I_c/2$, where the second term is the nonreciprocal component defined as $\Delta I_c = I_c^+ - I_c^-$. Figure 2c shows the $H$ dependence of $\Delta I_c$ in the temperature range from $T = 4.2$ to 11 K. Below $T = 9$ K, the antisymmetric behavior of $\Delta I_c$-$B$ curve is apparently observed, in that $|\Delta I_c|$ first linearly increases from zero with increasing (decreasing) $B$ up to 4 T (down to −4 T) and then suppressed by further increase (decrease) of $B$. The values of diode efficiency $\eta = (I_c^+ - I_c^-)/(I_c^+ + I_c^-) = \Delta I_c/2I_c^0$ at those temperatures are summarized in Fig. 2d. The antisymmetric $\Delta I_c(B)$ and $\eta(B)$ with peak and valley structures are common feature of the superconducting diode effect observed in a wide range of polar superconductors[13,17,18,21]. Although the theoretical analysis is indispensable for interpretation of the suppression of $\eta$ at high magnetic field and low temperature, it is natural that $\Delta I_c$ approaches zero at the normal state, passing through the maximum, due to the shrinkage of the order parameter with increasing $B$.

**Magnetic-field dependence of second harmonic resistance.**

Next, to quantitatively evaluate the coefficient $\gamma$ in Eq. (1), the first and second harmonic voltages were measured using the lock-in technique with a low excitation current. When ac current $I_{ac} = \sqrt{2}I_0 \sin \omega t = \sqrt{2}J_0 A \sin \omega t$ is applied to the channel ($A$: cross-sectional area of the channel, $J_0$: applied current density), Eq. (1) yields the output voltage as follows:

$$\sqrt{2}V_{xx} = R_{xx}I_{ac} = \sqrt{2}R^0_{xx}J_0 A \sin \omega t + \gamma R^0_{xx} B J_0^2 A \sin\left(2\omega t + \frac{\pi}{2}\right) + \text{constant}. \quad \text{Eq. (2)}$$

Since the measured first and second harmonic resistances correspond to $R^\omega_{xx} = R^0_{xx}$ and $R^{2\omega}_{xx} = \gamma R^0_{xx} J_0 B / \sqrt{2}$, respectively, $\gamma$ can be experimentally obtained as $\gamma = \sqrt{2} R^{2\omega}_{xx} / R^\omega_{xx} J_0 B$.

Figure 3a shows $R^\omega_{xx}(B)$ and $R^{2\omega}_{xx}(B)$ with $I_0 = 0.2$ mA from the normal state at $T = 20$ K (above $T_c^{on}$) down to the superconducting state at $T = 4.2$ K. Below $T_c^{on}$, $R^\omega_{xx}(B)$ shows a positive magnetoresistance (MR) corresponding to the suppression of superconductivity by $B$. Along with the emergence of positive MR in $R^\omega_{xx}(B)$, the antisymmetric $R^{2\omega}_{xx}(B)$ develops with peak and valley structures at high-field region, which resembles those reported in polar or chiral superconductors such as WS$_2$ nanotube (Ref. 9), gated MoS$_2$ (Refs. 10,11) and SrTiO$_3$ (Ref. 12) as well as superconducting heterostructures[13,15,21]. With further decreasing $T$, such antisymmetric $R_{xx}^{2\omega}(B)$ is suppressed in the full superconducting state. Figure 3b presents temperature profile of $\gamma(T)$ together with $R_{xx}^{\omega}(T)$ around thermal creep regime with $I_0$ = 0.2 mA at $B$ = 1 T to 15 T. The $\gamma(T)$ curve exhibits two distinct features: (I) $\gamma$ is greatly enhanced and exhibits the peak, which shifts towards lower temperature when $B$ increases in accordance with the decrease of $T_c(B)$. The peak value of $\gamma$ reaching 3.1 × $10^{-10}$ m$^2$A$^{-1}$T$^{-1}$ at $B$ = 5 T corresponds to $10^6$ times enhancement from the normal state (a black circle) as discussed later. (II) The value of $\gamma$ rapidly decreases with further decreasing $T$ toward zero resistance. This corresponds to the freezing of the vortices, resulting in the suppression of vortex motion at the present low excitation current. The concomitant occurrence of the enhancement of $\gamma$ and the $R_{xx}(T)$ drop described by TAFF model in $T_c^{\text{zero}} < T < T_c^{\text{on}}$ implicates that the activation energy (pinning potential) is effectively asymmetric against the polarity of current and magnetic field, which is the origin of the nonreciprocal resistance.

**Universality of supercurrent rectification within the vortex ratchet motion model.**

To link the driving mechanisms of two nonreciprocal phenomena at high and low excitation currents, we compare the values of $\eta$ and $\gamma$ as a function of $B$ and $T$. As seen in Fig. 2d, the value of $\eta(T)$ monotonically increases toward low temperature. In contrast, the value of $\gamma(T)$ is more dramatically enhanced than $\eta(T)$ with decreasing $T$ in the thermal creep region. Considering the different $T$ dependence between them, we empirically found a universal scaling law between $\gamma$ and $\eta/(BT)$ within the thermal creep regime as shown in the yellow shaded region of Fig. 3c. This linear scaling relation indicates that the rapid increase in $\gamma(T,B)$ is described as the increase in $\eta$ multiplied by $1/TB$. By using the rather high voltage criteria of 1 mV, it is possible to compare $\gamma$ and $\eta$ in the common region in the $B$-$T$ plane, obtained by the independent experiments with low and high excitation current, respectively.

To interpret the scaling law found in this work, we now consider a simple model assuming the asymmetric pinning potential $U_0(z)$, as schematically shown in the inset of Fig. 3c, which leads to the asymmetric creep or depinning of vortices. The model based on the depinning process is feasible from the fact that the measured critical current density $J_c$ is much smaller than the depairing critical current $J_d$ (Refs. 42,43). As mentioned above, in the TAFF scenario, the nonreciprocal contribution to $R_{xx}$ below $T_c^{on}$ originates from the activated motion of vortices in asymmetric pinning potential, which is referred to vortex ratchet effect[11]. In such condition, the depinning critical current densities $J_c^+$ and $J_c^-$ are calculated by balancing the Lorentz forces, $J_c^+ B V_c$ and $J_c^- B V_c$, working in the upward- and the downward-directions along the $z$ axis, and the pinning forces, $U_0/d_{up}$ and $U_0/d_{down}$ in the opposite directions to the Lorentz forces, respectively. Here, $d_{up}$ and $d_{down}$ represent the sizes of $U_0(T,B)$ from its minimum point with $d_{up} + d_{down} = 2d_0$ and $V_c$ the effective volume of vortex bundle. Then, we obtain,

$$d_{up} = (1+\eta)d_0, \qquad \text{Eq. (3a)}$$

$$d_{down} = (1-\eta)d_0, \qquad \text{Eq. (3b)}$$

from $\eta = \Delta I_c / 2 I_c^0$. By applying Eqs. (3a) and (3b) in the TAFF model, the emergent $\gamma$ as a function of $\eta$ in the low excitation current is derived as,

$$\gamma = \frac{V_c d_0}{k_B} \frac{\eta}{T}, \qquad \text{Eq. (4)}$$

where $k_B$ is the Boltzmann constant. In the case that the vortices move individually like in the TAFF regime, $V_c$ can be described as $\lambda a^2 = \lambda \left(\frac{\Phi_0}{B}\right)$ with $\lambda$, $a \approx \sqrt{\frac{\Phi_0}{B}}$ and $\Phi_0$ being a penetration depth, the average distance between vortices and the flux quantum $h/2e$ with $h$ the Planck constant, respectively. The substitution of it to $V_c$ in Eq. (4) leads to $\gamma = \frac{\Phi_0 \lambda d_0}{k_B} \frac{\eta}{BT}$, which explains the empirical scaling law found in Fig. 3c. The linear relation of $\gamma$-$\eta/BT$ yields a size of the pinning potential $d_0 \sim 2.3$ nm by assuming the penetration depth $\lambda \sim 580$ nm in our sample. The comparable value of $d_0$ to the superconducting coherence lengths ($\xi_{ab}$ = 2.2 nm and $\xi_c$ = 1.2 nm) ensures the validity of our model.

**Origin of the asymmetric pinning potential.**

Finally, we discuss the possible origins of the asymmetric pinning potential in the Se/Fe(Se,Te)/FeTe heterostructure. Figure 4a and 4b show $R_{xx}^{2\omega}(B)$ with different $I_0$ in the thermal creep regime at $T$ = 10 K and in the normal state at $T$ = 20 K, respectively. In addition to the enhanced $R_{xx}^{2\omega}(B)$ with antisymmetric peak and valley structures in the thermal creep regime (Fig. 4a), we observed a linear $H$-dependence of the $R_{xx}^{2\omega}(B)$ in the normal state (Fig. 4b). This non-zero nonreciprocal response in the normal state is suggestive of the strong effect of the spin-orbit interaction[6,7] as previously exemplified in the polar conductors[8,12] and the topological insulator heterostructures[14,15]. Indeed, as shown in Fig. 4c, the slope of $R_{xx}^{2\omega}(B)$, $R_{xx}^{2\omega}/B$, in the normal state (blue circles), is linearly dependent on $I_0$ (dashed line in Fig. 4c), being consistent with Eq. (1). From the linear relation, we extract the normal-state nonreciprocal coefficient of $\gamma$ = 1.1 × 10$^{-16}$ m$^2$A$^{-1}$T$^{-1}$. It is noted that the strong enhancement of $R_{xx}^{2\omega}/B$ (black dashed line in Fig. 4a) in the thermal creep regime by a factor of almost 10$^4$ as shown in Fig. 4c, originates from the enhanced $\gamma$ value by ~10$^6$ times due to the TAFF (Fig. 3b).

By considering the observation of nonreciprocal resistance even in the normal state, it is natural that the presence of the in-plane helical spin-momentum-locked state due to the Rashba effect[6,7,8] should play a role in the formation of asymmetric pinning potential. In such a state, the Cooper pairs in the superconducting state acquire a non-zero center-of-mass momentum $\Delta\boldsymbol{q}$ along *x* direction when **B** is applied along *y* direction as shown in Fig. 4d. This finite Cooper pair's momentum $\Delta\boldsymbol{q}$ may cause asymmetricity of superconducting order parameters with respect to the polarity of $\Delta\boldsymbol{q}$. According to the theoretical prediction assuming the external current **I** applied along $\Delta\boldsymbol{q}$, the condensation energy becomes asymmetric against the polarity of **I**, leading to the nonreciprocal $J_d(B)$ (Ref. 28). The same argument can be applied to the rotational supercurrent $\boldsymbol{J}_v$ in the absence of external current. Since the $\boldsymbol{J}_v$ has both components parallel and antiparallel to $\Delta\boldsymbol{q}$, depending on the radial position around the vortex core, the spatial distribution of the condensation energy density $E_{SC}(z)$ around the vortex core can be asymmetric with respect to the direction perpendicular to $\Delta\boldsymbol{q}$, which is parallel to the Lorentz force (*z* direction). Therefore, the asymmetric pinning potential, which is proportional to $E_{SC}(z)$, can be realized as a consequence of Rashba superconductors. The important correlation between vortex dynamics and spin-orbit interaction is found from the universal scaling law of the superconducting rectification effect, which will stimulate further theoretical studies.

In summary, we demonstrated that the simple vortex ratchet model incorporating the asymmetric vortex depinning phenomena can clearly explain the relationship between superconducting diode effect and nonreciprocal resistance. Our findings will open the perspectives of structural design not only for further functionalization of superconducting diode effect devices, but also for exploring exotic superconductors with spontaneously-broken time-reversal symmetry.

**Materials and Methods**

The Se-capped 23.3-nm-thick Fe(Se,Te)/20.0-nm-thick FeTe heterostructure and 57.1-nm-thick FeTe single layer were prepared on CaF$_2$ substrates by pulsed-laser deposition at growth at 350°C in vacuum. The sharp interface without noticeable interdiffusion were confirmed by structural characterizations using x-ray diffraction and scanning transmission electron microscopy (STEM) with composition mapping by energy-dispersive x-ray spectroscopy images. Thickness of each layer of the heterostructure was calibrated from deposition rates of the individual Fe(Se,Te) and FeTe single-layer thin-film samples and STEM image. The Se/Fe(Se,Te)/FeTe heterostructure was patterned into a rectangular-shaped device with center-to-center distance of voltage probes and channel width being $L = 155$ and $W = 50$ $\mu$m for precise measurement of current-voltage (*I-V*) characteristics by water lift-off technique. The current-voltage (*I-V*) characteristics were obtained by sweeping DC current, resulting in the normal-state resistance evaluated as $R_{IV}^{N} = V/I = 272$ $\Omega$ at the current much higher than $I_c$ and $T = 4.2$ K. This value is consistent with the four-terminal resistance $(L/W)R_{xx}^{N} = 273$ $\Omega$ at $T = 20$ K measured with 10 $\mu$A, ensuring that the effect of Joule heating is excluded. The sheet resistance and its nonreciprocal contribution were characterized by first and second harmonic resistance measurements using lock-in technique at a frequency of 13 Hz. All electrical measurements were performed in a variable temperature insert equipped with a 15 T superconducting magnet.

**Reference**

1. Reyren, N., Thiel, S., Caviglia, A. D., Fitting Kourkoutis, L., Hammerl, G., Richter, C., Schneider, C. W., Kopp, T., Rüetschi, A.-S., Jaccard, D., Gabay, M., Muller, D. A., Triscone, J.-M., & Mannhart, J., Superconducting Interfaces Between Insulating Oxides. Science **317**, 1196-1199 (2007).

2. Kozuka, Y., Kim, M., Bell, C., Kim, B. G., Hikita, Y., & Hwang, H. Y., Two-dimensional normal-state quantum oscillations in a superconducting heterostructure. Nature **462**, 487-490 (2009).

3. He, Q. L., Liu, H., He, M., Lai, Y. H., He, H., Wang, G., Law, K. T., Lortz, R., Wang, J., & Sou, I. K., Two-dimensional superconductivity at the interface of a $Bi_2Te_3$/FeTe heterostructure. Nature Commun. **5**, 4247 (2014).

4. Cao, Y., Fatemi, V., Fang, S., Watanabe, K., Taniguchi, T., Kaxiras, E., & Jarillo-Herrero, P., Unconventional superconductivity in magic-angle graphene superlattices. Nature **556**, 43-50 (2018).

5. Lin, J. X., Siriviboon, P., Scammell, H. D., Liu, S., Rhodes, D., Watanabe, K., Taniguchi, T., Hone, J., Scheurer, M. S., & Li, J. I. A., Zero-field superconducting diode effect in small-twist-angle trilayer graphene. Nature Phys. **18**, 1221-1227 (2022).

6. Tokura, Y., & Nagaosa, N., Nonreciprocal responses from noncentrosymmetric quantum materials. Nature Commun. **9**, 3740 (2018).

7. Ideue, T., & Iwasa, Y., Symmetry Breaking and Nonlinear Electric Transport in van der Waals Nanostructures. Annu. Rev. Condens. Matter Phys. **12**, 201-223 (2021).

8. Ideue, T., Hamamoto, K., Koshikawa, S., Ezawa, M., Shimizu, S., Kaneko, Y., Tokura, Y., Nagaosa, N., & Iwasa, Y., Bulk rectification effect in a polar semiconductor. Nature Phys. **13**, 578-583 (2017).

9. Qin, F., Shi, W., Ideue, T., Yoshida, M., Zak, A., Tenne, R., Kikitsu, T., Inoue, D., Hashizume, D., & Iwasa, Y., Superconductivity in a chiral nanotube. Nature Commun. **8**, 14465 (2017).

10. Wakatsuki, R., Saito, Y., Hoshino, S., Itahashi, Y. M., Ideue, T., Ezawa, M., Iwasa, Y., & Nagaosa, N., Nonreciprocal charge transport in noncentrosymmetric superconductors. Sci. Adv. **3**, e1602390 (2017).

11. Itahashi, Y. M., Saito, Y., Ideue, T., Nojima, T., & Iwasa, Y., Quantum and classical ratchet motions of vortices in a two-dimensional trigonal superconductor. Phys. Rev. Research **2**, 023127 (2020).

12. Itahashi, Y. M., Ideue, T., Saito, Y., Shimizu, S., Ouchi, T., Nojima, T., & Iwasa, Y., Nonreciprocal transport in gate-induced polar superconductor $SrTiO_3$. Sci. Adv. **6**,

eaay9120 (2020).

13. Lustikova, J., Shiomi, Y., Yokoi, N., Kabeya, N., Kimura, N., Ienaga, K., Kaneko, S., Okuma, S., Takahashi, S., & Saitoh, E., Vortex rectenna powered by environmental fluctuations. Nat. Commun. **9**, 4922 (2018).

14. Yasuda, K., Yasuda, H., Liang, T., Yoshimi, R., Tsukazaki, A., Takahashi, K. S., Nagaosa, N., Kawasaki, M., & Tokura, Y., Nonreciprocal charge transport at topological insulator/superconductor interface. Nature Commun. **10**, 2734 (2019).

15. Masuko, M., Kawamura, M., Yoshimi, R., Hirayama, M., Ikeda, Y., Watanabe, R., He, J. J., Maryenko, D., Tsukazaki, A., Takahashi, K. S., Kawasaki, M., Nagaosa, N., & Tokura, Y., Nonreciprocal charge transport in topological superconductor candidate $Bi_2Te_3$/$PdTe_2$ heterostructure. npj Quantum Mater. **7**, 104 (2022).

16. Hou, Y., Nichele, F., Chi, H., Lodesani, A., Wu, Y., Ritter, M. F., Haxell, D. Z., Davydova, M., Ilić, S., Glezakou-Elbert, O., Varambally, A., Bergeret, F. S., Kamra, A., Fu, L., Lee, P. A., & Moodera, J. S., Ubiquitous Superconducting Diode Effect in Superconductor Thin Films. Phys. Rev. Lett. **131**, 027001 (2023).

17. Bauriedl, L., Bäuml, C., Fuchs, L., Baumgartner, C., Paulik, N., Bauer, J. M., Lin, K. Q., Lupton, J. M., Taniguchi, T., Watanabe, K., Strunk, C., & Paradiso, N., Supercurrent diode effect and magnetochiral anisotropy in few-layer $NbSe_2$. Nature

Commun. **13**, 4266 (2022).

18. Suri, D., Kamra, A., Meier, T. N. G., Kronseder, M., Belzig, W., Back, C. H., & Strunk, C., Non-reciprocity of vortex-limited critical current in conventional superconducting micro-bridges. Appl. Phys. Lett. **121**, 102601 (2022).

19. Rikken, G. L. J. A., Fölling, J., & Wyder, P., Electrical Magnetochiral Anisotropy. Phys. Rev. Lett. **87**, 236602 (2001).

20. Rikken, G. L. J. A., & Wyder, P., Magnetoelectric Anisotropy in Diffusive Transport. Phys. Rev. Lett. **94**, 016601 (2005).

21. Ando, F., Miyasaka, Y., Li, T., Ishizuka, J., Arakawa, T., Shiota, Y., Moriyama, T., Yanase, Y., & Ono, T., Observation of superconducting diode effect. Nature **584**, 373-376 (2020).

22. Narita, H., Ishizuka, J., Kawarazaki, R., Kan, D., Shiota, Y., Moriyama, T., Shimakawa, Y., Ognev, A. V., Samardak, A. S., Yanase, Y., & Ono, T., Field-free superconducting diode effect in noncentrosymmetric superconductor/ferromagnet multilayers. Nature Nanotech. **17**, 823-828 (2022).

23. Vodolazov, D. Y., & Peeters, F. M., Superconducting rectifier based on the asymmetric surface barrier effect. Phys. Rev. B **72**, 172508 (2005).

24. Villegas, J. E., Savel'ev, S., Nori, F., Gonzalez, E. M., Anguita, J. V., García, R., &

Vicent, J. L., A Superconducting Reversible Rectifier That Controls the Motion of Magnetic Flux Quanta. Science **302**, 1188-1191 (2003).

25. Villegas, J. E., Gonzalez, E. M., Gonzalez, M. P., Anguita, J. V., & Vicent, J. L., Experimental ratchet effect in superconducting films with periodic arrays of asymmetric potentials. Phys. Rev. B **71**, 024519 (2005).

26. de Souza Silva, C. C., Van de Vondel, J., Morelle, M., & Moshchalkov, V. V., Controlled multiple reversals of a ratchet effect. Nature **440**, 651-654 (2006).

27. Lyu, Y. Y., Jiang, J., Wang, Y. L., Xiao, Z. L., Dong, S., Chen, Q. H., Milošević, M. V., Wang, H., Divan, R., Pearson, J. E., Wu, P., Peeters, F. M., & Kwok, W. K., Superconducting diode effect via conformal-mapped nanoholes. Nature Commun. **12**, 2703 (2021).

28. Daido, A., Ikeda, Y., & Yanase, Y., Intrinsic Superconducting Diode Effect, Phys. Rev. Lett. **128**, 037001 (2022).

29. Yuan, N. F. Q., & Fu, L., A Supercurrent diode effect and finite-momentum superconductors. Proc. Natl. Acad. Sci. U. S. A. **119** e2119548119 (2022).

30. Hoshino, S., Wakatsuki, R., Hamamoto, K., & Nagaosa, N., Nonreciprocal charge transport in two-dimensional noncentrosymmetric superconductors. Phys. Rev. B **98**, 054510 (2018).

31. Gutfreund, A., Matsuki, H., Plastovets, V., Noah, A., Gorzawski, L., Fridman, N., Yang, G., Buzdin, A., Millo, O., Robinson, J. W. A., & Anahory, Y., Direct observation of a superconducting vortex diode. Nature Commun. **14**, 1630 (2023).

32. Hsu, F. C., Luo, J. Y., Yeh, K. W., Chen, T. K., Huang, T. W., Wu, P. M., Lee, Y. C., Huang, Y. L., Chu, Y. Y., Yan, D. C., & Wu, M. K., Superconductivity in the PbO-type structure $\alpha$-FeSe. Proc. Natl. Acad. Sci. USA **105**, 14262-14264 (2008).

33. Fernandes, R. M., Coldea, A. I., Ding, H., Fisher, I. R., Hirschfeld, P. J., & Kotliar, G., Iron pnictides and chalcogenides: a new paradigm for superconductivity. Nature **601**, 35-44 (2022).

34. Lei, H., Hu, R., & Petrovic, C., Critical fields, thermally activated transport, and critical current density of $\beta$-FeSe single crystals. Phys. Rev. B **84**, 014520 (2011).

35. Imai, Y., Sawada, Y., Nabeshima, F., & Maeda, A., Suppression of phase separation and giant enhancement of superconducting transition temperature in FeSe$_{1-x}$Te$_x$ thin films. Proc. Natl. Acad. Sci. USA **112**, 1937-1940 (2015).

36. Ishida, K., Onishi, Y., Tsujii, M., Mukasa, K., Qiu, M., Saito, M., Sugimura, Y., Matsuura, K., Mizukami, Y., Hashimoto, K., & Shibauchi, T., Pure nematic quantum critical point accompanied by a superconducting dome. Proc. Natl. Acad. Sci. USA **119**, e2110501119 (2022).

37. Lei, H., Hu, R., Choi, E. S., Warren, J. B., & Petrovic, C., Pauli-limited upper critical field of $Fe_{1+y}Te_{1-x}Se_x$. Phys. Rev. B **81**, 094518 (2010).

38. Si, W., Han, S. J., Shi, X., Ehrlich, S. N., Jaroszynski, J., Goyal, A., & Li, Q., High current superconductivity in $FeSe_{0.5}Te_{0.5}$-coated conductors at 30 tesla. Nature Commun. **4**, 1347 (2013).

39. Zhang, P., Yaji, K., Hashimoto, T., Ota, Y., Kondo, T., Okazaki, K., Wang, Z., Wen, J., Gu, G. D., Ding, H., & Shin, S., Observation of topological superconductivity on the surface of an iron-based superconductor. Science **360**, 182-186 (2018).

40. Palstra, T. T. M., Batlogg, B., van Dover, R. B., Schneemeyer, L. F., & Waszczak, J. V., Dissipative flux motion in high-temperature superconductors. Phys. Rev. B **41**, 6621 (1990).

41. Kwok, W. K., Olsson, R. J., Karapetrov, G., Welp, U., Vlasko-Vlasov, V., Kadowaki, K., & Crabtree, G. W., Modification of vortex behavior through heavy ion lithography. Physica C: Superconductivity **382**, 137-141 (2002).

42. Shmidt, V. V., Critical currents in superconductors, Sov. Phys. Usp. **13**, 408 (1970).

43. Shmidt, V. V., The critical current in superconducting films. Sov. Phys. JETP **30**, 1137-1142 (1970).

**Acknowledgments**

The authors thank Shun Ito and Kana Takenaka at Analytical Research Core for Advanced Materials, Institute for Materials Research, Tohoku University, for cross-sectional scanning transmission electron microscope observation, and Terukazu Nishizaki and Atsushi Tsukazaki for stimulating discussion. This work was supported by JST, PRESTO Grant No. JPMJPR21A8, JSPS KAKENHI Grant Nos. JP23K26379 and JP21K18889, Iketani Science and Technology Foundation, and Tanikawa Foundation.

**Figure captions**

**Fig. 1 | Superconducting properties of a Fe(Se,Te)/FeTe heterostructure. a,** Schematics of Fe(Se,Te)/FeTe heterostructure and superconducting diode effect in the presence of quantum vortices. **b,** $R_{xx}$-$T$ characteristic of the Fe(Se,Te)/FeTe heterostructure (red solid line) and the reference 57.1-nm-thick FeTe single layer (green solid line). (Inset) $R_{xx}$-$T$ curve around superconducting transition. **c,** Superconducting critical parameters of the Fe(Se,Te)/FeTe heterostructure (red) and critical surface of bulk FeSe single crystal[32,34] (black circles). The basal plane shows normalized resistance, $R_{xx}/R_N$ where $R_N$ is normal-state resistance at 20 K in the $B$-$T$ plane, and vortex phase diagram.

**Fig. 2 | Superconducting diode effect. a,** Current-voltage (*I-V*) characteristics when the in-plane magnetic field of $\pm 5$ T perpendicular to the current is applied at $T$ = 4.2 K. Red and black solid lines indicate the positive and negative polarity of the magnetic field. **b,** The magnetic field dependence of critical current for positive ($I_c^+$, red) and negative ($I_c^-$, blue) direction. Black solid line indicates the averaged value defined as $I_c^0 = (I_c^+ + I_c^-)/2$. **c,d,** The magnetic-field dependence of (**c**) $\Delta I_c = I_c^+ - I_c^-$ and (**d**) rectification efficiency $\eta = \Delta I_c / 2 I_c^0$ for $T$ = 4.2 (red), 7.0 (orange), 9.0 (green), and 11.0

K (blue). For clarity, the data are offset in **c**.

**Fig. 3 | Enhancement of the nonreciprocal coefficient in the thermal creep regime and universal behavior. a,** Magnetic-field dependence of (top) first and (bottom) second harmonic resistances ($R_{xx}^{\omega}$ and $R_{xx}^{2\omega}$, respectively) at $T$ = 4.2, 7, 9, 10, 12, and 20 K for the ac current $I_0$ = 0.2 mA. For clarity, $R_{xx}^{2\omega}$ is magnified by 10 at $T$ = 4.2 and 7 K and by $10^2$ at $T$ = 20 K. **b,** The $R_{xx}^{\omega}(T)$ and $\gamma$ curves by sweeping the temperature measured at $I_0$ = 0.2 mA for $B$ from 1 (red) to 15 T (purple) with an interval of 2 T. A black circle indicates the $\gamma$ value at normal state ($T$ = 20 K). **c,** The relationship between $\gamma$ and $\eta/BT$ obtained for magnetic field $B$ = 1 to 15 T. (Inset) Schematic of asymmetric pinning potential around the vortex core.

**Fig. 4 | The origin of the asymmetric pinning potential of vortices. a,b,** Magnetic-field dependence of $R_{xx}^{2\omega}$ for (**a**) thermal creep regime at $T$ = 10 K and (**b**) normal state at $T$ = 20 K with different $I_0$. **c,** $I_0$ dependence of the slope of $R_{xx}^{2\omega}(H)$ for thermal creep regime (red) and normal state (blue). For thermal creep, the slope of $R_{xx}^{2\omega}(H)$ was obtained around low field region (black dashed line in **a**). **d,** Schematics of helical spin structure under in-plane $\boldsymbol{B}$, rotational current $\boldsymbol{J}_v$ around a vortex, and asymmetric condensation

energy density $E_{\text{sc}}(z)$.

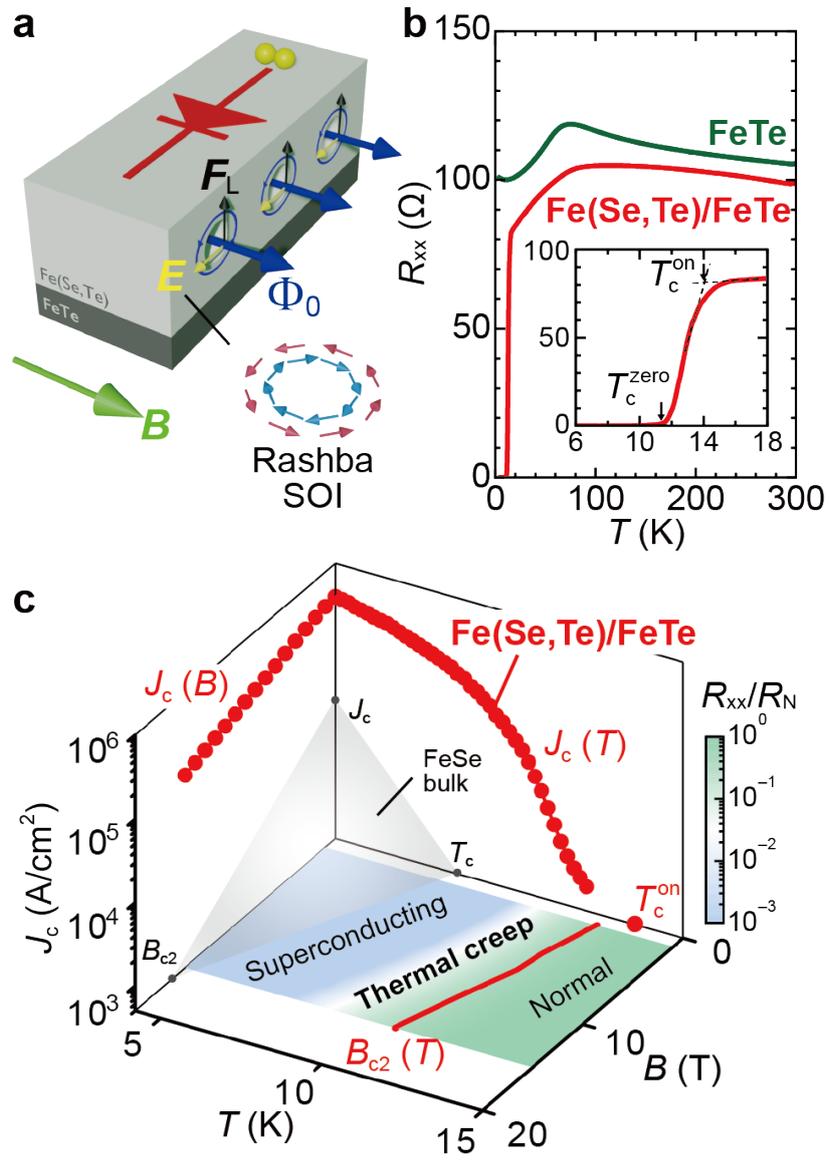

Figure 1 Kobayashi *et al*

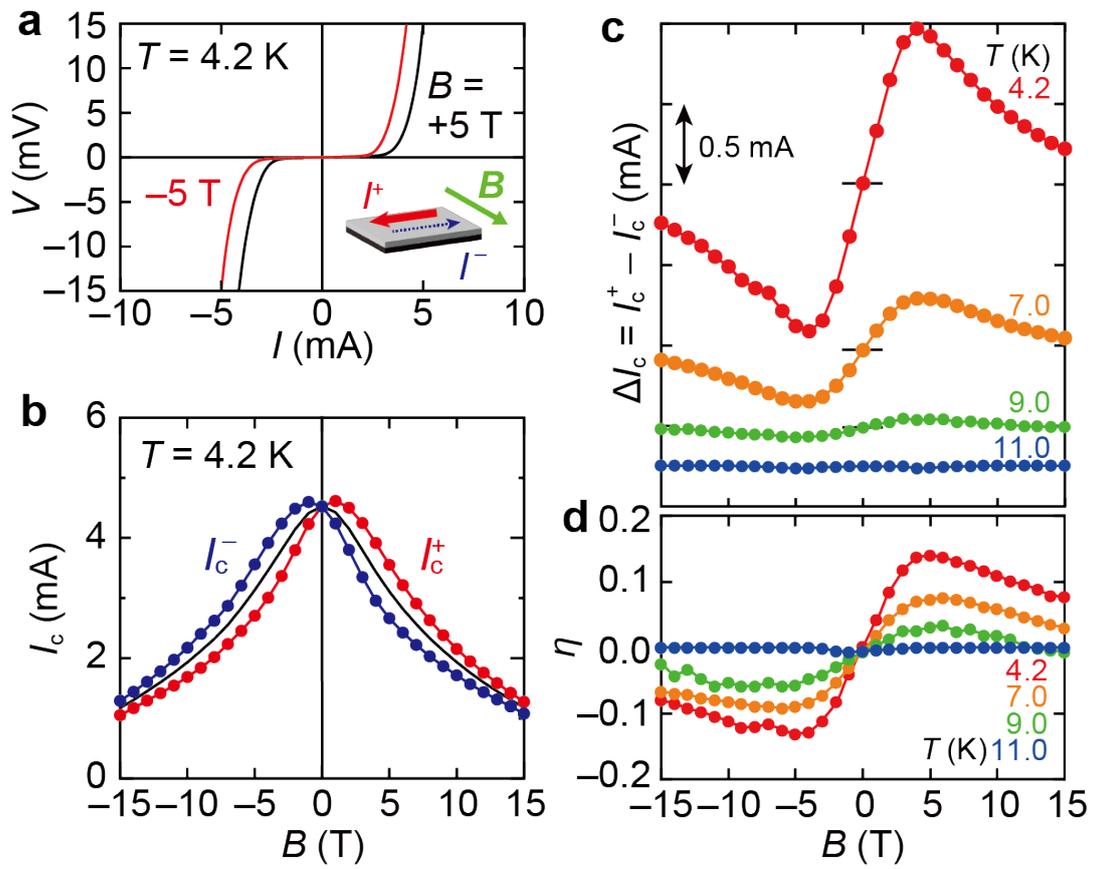

Figure 2 Kobayashi *et al*

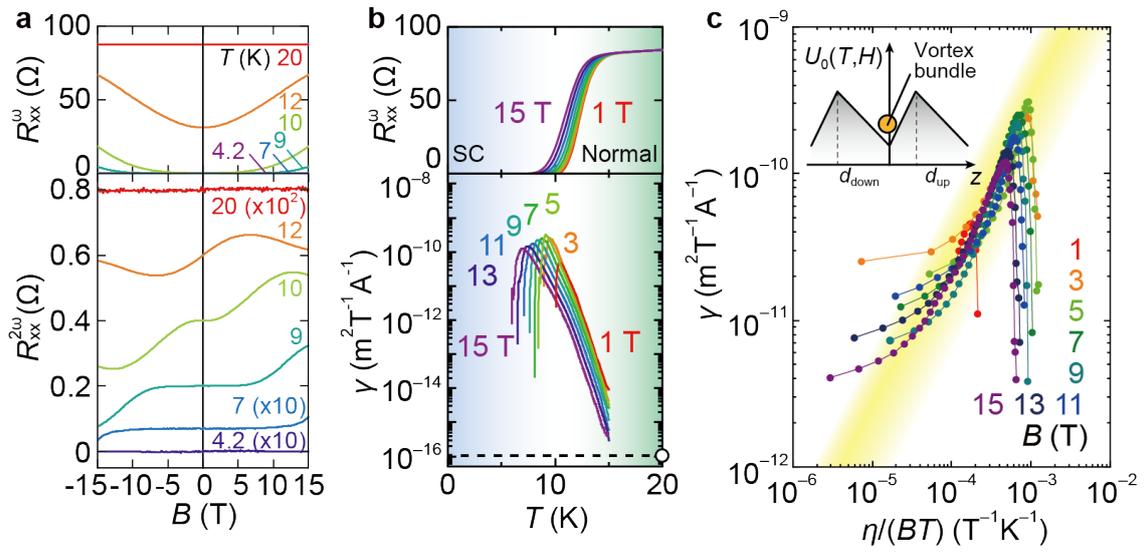

Figure 3 Kobayashi *et al*

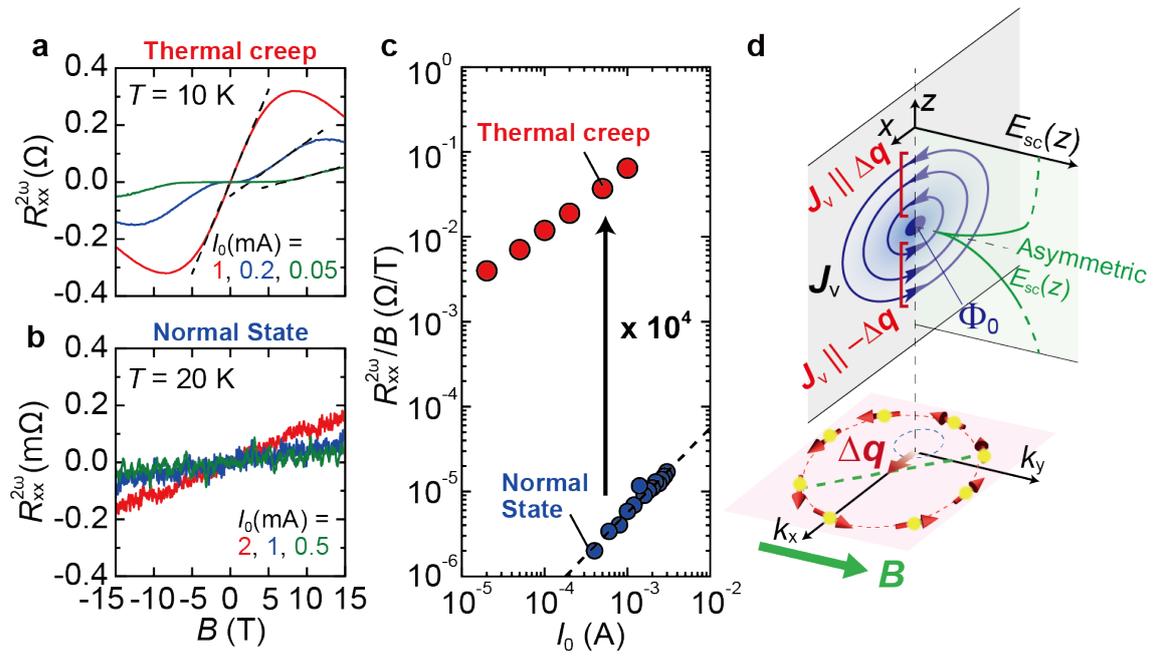

Figure 4 Kobayashi *et al*